\documentclass[a4paper,aps,preprint,nofootinbib]{revtex4}
\pdfoutput=1
\usepackage[latin1]{inputenc}
\usepackage{bbm}
\usepackage{bm}
\usepackage{graphicx}
\usepackage{epsfig}
\usepackage{subfigure}
\usepackage{latexsym}
\usepackage{amsmath}
\usepackage{amsfonts}
\usepackage{amssymb}
\usepackage{wasysym}
\usepackage{color}
\usepackage{slashed}

\newcommand{\be}{\begin{equation}}
\newcommand{\ee}{\end{equation}}

\newcommand{\bea}{\begin{eqnarray}}
\newcommand{\eea}{\end{eqnarray}}

\definecolor{rossoCP3}{cmyk}{0,.88,.77,.40}
\begin{document}


\title{Vector resonances at LHC Run II\\ in composite 2HDM}
\author{Stefano Di Chiara$ ^{a}$\footnote{stefano.dichiara@kbfi.ee}}
\author{Matti Heikinheimo$ ^{b,c}$\footnote{matti.heikinheimo@helsinki.fi}}
\author{Kimmo Tuominen $ ^{b,c}$\footnote{kimmo.i.tuominen@helsinki.fi}}
\affiliation{\mbox{ $ ^{a}$ National Institute of Chemical Physics and Biophysics, R\"avala 10, 10143 Tallinn, Estonia}}
\affiliation{\mbox{ $ ^{b}$ Helsinki Institute of Physics, P.O.Box 64, FI-000140, Univ. of Helsinki, Finland}}
\affiliation{\mbox{ $ ^{c}$ Department of Physics, P.O.Box 64, FI-000140, Univ. of Helsinki, Finland}}


\begin{abstract}
We consider a model where the electroweak symmetry breaking is driven by strong dynamics,
resulting in an electroweak doublet scalar condensate, and transmitted to the standard model
matter fields via another electroweak doublet scalar. At low energies the effective theory
therefore shares features with a type-I two Higgs doublet model. However, important differences arise
due to the rich composite spectrum expected to contain new vector resonances accessible
at the LHC. We carry out a systematic analysis of the vector resonance signals at LHC and find that the model remains viable, but will be tightly constrained by direct searches as
the projected integrated luminosity, around 200 fb$^{-1}$, of the current run becomes available.
\end{abstract}
\preprint{HIP-2016-36/TH}
\maketitle

\section{Introduction}
\label{Intro}
Recently both the ATLAS and CMS collaborations updated the direct search limits
on heavy vector resonances by using 13~TeV data with integrated luminosity ranging from
12.4 to 15.5 fb${^{-1}}$~\cite{ATLAS:2016cyf,ATLAS:2016ecs,ATLAS:2016yqq,ATLAS:2016cwq,ATLAS:2016npe,CMS:2016pfl,CMS:2016abv}.
The resulting lower limits on sequential $W'$
and $Z'$ boson masses are about 4.7 and 4~TeV,
respectively. The LHC reach for such heavy vector resonances is expected to
further increase in the next analyses given that the integrated
luminosity delivered by the LHC has reached about 45 fb${}^{-1}$ at 13~TeV at the end of the 2016 run.

A traditional class of models requiring the existence of heavy vector resonances
is technicolor (TC): the strong interaction responsible for a technifermion
condensate breaking the electroweak (EW) symmetry generates also a rich spectrum of
composite states whose mass is roughly fixed in the TeV range by the need to
provide the observed masses to the $W^\pm$ and $Z$ gauge bosons. The observation of TeV scale vector resonances at LHC would therefore be a
strong hint that technicolor is the underlying theory realized in Nature.
The current mass limits are comparable to the typical TC scale, equal to about
$4\pi v_{w}\simeq 3$~TeV, hence it is important to test the viability of TC
theories and determine what portion of parameter space is soon to be
explored, and whether a negative outcome of a heavy vector search could in principle
rule out some of the currently viable theories in the TC framework.

In TC the Higgs couplings, which are constrained by LHC
measurements within a few percent uncertainty, depend on the particular
ultraviolet (UV) completion used to transmit EW symmetry breaking to the
Standard Model (SM) fermion sector. To simplify the phenomenological analysis and its comparison with LHC data
we choose a simple TC model as a template for more general extended TC
theories. In the TC model at hand the interactions between the technifermions
and SM fermions are mediated by an EW doublet scalar
field~\cite{Simmons:1988fu,Dine:1990jd,Samuel:1990dq,Kagan:1990az,Carone:1992rh,Carone:1994mx}
(which we treat as elementary but can in principle be composite as well).
The scalar sector of such UV complete model corresponds to that of a composite
two Higgs doublet model\footnote{Such name has been used originally in the context of
composite pseudo-Goldstone boson Higgs models~\cite{Gripaios:2009pe}.}
(2HDM)~\cite{Zerwekh:2009yu,Antola:2009wq,Alanne:2013dra,DeCurtis:2016scv,DeCurtis:2016tsm,Agugliaro:2016clv,Belyaev:2016ftv}.
Due to the strong interacting dynamics the model features a spectrum of higher-spin composite states, which distinguish this model from the ordinary type-I 2HDM. The main goal of this paper is to test the viability of this model given the LHC measurements of the light Higgs couplings to SM particles and LHC direct search constraints on heavy scalar and vector resonances, and to determine the expected reach of LHC Run II for the vector heavy states appearing in this model.

The paper is organized as follows: In section \ref{model} we will first briefly introduce the model.
In section \ref{couplingsfit} we update the fit of the model parameters to concur with the LHC data on the Higgs couplings, and
in section \ref{LHCsigs} we confront the signals of the vector resonances of the model with the
current direct search constraints and present projections for the near future reach of the LHC experiments. Then in section
\ref{checkout} we offer our conclusions and a brief outlook for future research.

\section{Composite Vector Boson Interactions}
\label{model}
The model we study, a composite 2HDM~\cite{Antola:2009wq,Alanne:2013dra},
extends the particle content of the Next to Minimal Walking Technicolor (NMWT)
\cite{Sannino:2004qp, Dietrich:2005jn, Belyaev:2008yj}
by a scalar field, $H$ (to be considered as a
remnant of a UV complete theory, not necessarily strongly interacting), which
features the same couplings as the SM Higgs field, and moreover couples to the
technifermion fields via a renormalizable Yukawa interaction with coupling $y_{\rm TC}$.
At an energy below the scale of the TC strong interaction,
$\Lambda_{\rm TC}\sim4\pi v_{w}\simeq 3$~TeV, the strongly interacting technifermions
form a tower of composite states, analogously to QCD, whose interactions are
encoded in an effective Lagrangian featuring the same global symmetries as the
fundamental theory. The scalar sector of the 2HDM effective Lagrangian is
expressed in terms of $H$ and a composite matrix scalar field $M$ as follows
\bea\label{Ls}
{\cal L}_{scalar} &=& \frac{1}{2} {\rm Tr}\!\left(D_\mu H^\dagger D^\mu H+D_\mu
M^\dagger D^\mu M-m_H^2 H^\dagger H- m_M^2 M^\dagger M\right) \nonumber \\
&+&\left[\frac{y_{\rm TC}}{2}  {\rm Tr}\!\left(c_3 D_\mu M^\dagger D^\mu H+c_1 f^2
M^\dagger H\right)+\frac{c_2y_{\rm TC}}{24} {\rm Tr}(M^\dagger M)
{\rm Tr}(M^\dagger H)\right. \nonumber \\
&+&\left.\frac{c_4 y_{\rm TC}}{24}\lambda_H{\rm Tr}(H^\dagger H){\rm Tr}(M^\dagger H)+{\textrm{h.c.}}\right]-\frac{\lambda_H}{24}{\rm Tr}\!\left(H^\dagger H\right)^2
-\frac{\lambda_M}{24}{\rm Tr}\!\left(M^\dagger M\right)^2,
\eea
with
\be\label{HM}
H=\frac{1}{\sqrt{2}}\left( \phi I_{2\times 2}+i \pi_H^k\sigma_k \right)\,,\ \langle h\rangle=v\,,\quad M=\frac{1}{\sqrt{2}}\left( \varphi I_{2\times 2}+i \pi_M^k\sigma_k \right)\,,\ \langle s\rangle=f\,,\quad
\ee
and
\be\label{covD}
D^\mu H=\partial^\mu H-i g_L W_a^\mu T^a H+i g_Y H\sigma_3  \quad,\
D^\mu M=\partial^\mu M-i g_L W_a^\mu T^a M+i g_Y M\sigma_3\,,
\ee
where $\sigma_k$ are Pauli matrices. Contrary to the SM case, in composite 2HDM
the squared mass term of the Higgs field $H$ is assumed to be positive
(and generally of the order of the physical Higgs mass), so that EW symmetry
breaking is triggered by the techniquark condensate $M$, with the EW scale determined in terms of the vevs of $M$ and $H$ by
\be\label{vfvw}
v_w^2=v^2+f^2+2 c_3 y_{TC} f v=(246\ {\rm GeV})^2\,.
\ee
At lower order in
$y_{\rm TC}$, considered to be perturbatively small, the composite vectors
$A_L^\mu$ and $A_R^\mu$ couple only to the field $M$~\cite{Alanne:2013dra},
\bea\label{Lv}
{\cal L}_{vector}&=&-\frac{1}{2}{\rm Tr}\!\left(\tilde{W}^{\mu\nu}\tilde{W}_{\mu\nu}\right)-\frac{1}{4}\tilde{B}^{\mu\nu}\tilde{B}_{\mu\nu}-\frac{1}{2}{\rm Tr}\!\left(F_L^{\mu\nu}F_{L\mu\nu}+F_R^{\mu\nu}F_{R\mu\nu}\right)+m_A^2 {\rm Tr}\!\left(C_{L\mu}^2+C_{R\mu}^2\right)\nonumber\\
&&-g^2_{\rm TC} r_2 {\rm Tr}\!\left(C_{L\mu} M C^\mu_R M^{\dagger}\right)+\frac{g_{\rm TC}^2 r_1}{4}{\rm Tr}\!\left(C_{L\mu}^2+C_{R\mu}^2\right) {\rm Tr}\! \left(M^\dagger M\right)\ ,
\eea
with
\be
C_L^\mu=A_L^\mu-\frac{g_L}{g_{\rm TC}}\tilde{W}^\mu\ ,\quad C_R^\mu=A_R^\mu-\frac{g_Y}{g_{\rm TC}} \tilde{B}^\mu\,,
\label{cfactors}
\ee
where $\tilde{W}$ and $\tilde{B}$ are the SU$(2)_L$ and U$(1)_Y$ gauge fields,
respectively. In Eq.~\eqref{Lv} we neglected derivative couplings of the
composite vectors given that these are anyway constrained to be small by the measured small values of the
oblique parameters \cite{Peskin:1991sw,Agashe:2014kda}. The fermion sector is the same as
that of the SM. The full Lagrangian, as defined in Eqs.~(\ref{Ls})-(\ref{cfactors}),
features the global symmetry SU$(2)_L\times$SU$(2)_R$,
broken by the bosonic $y_{\rm TC}$
couplings in Eq.~\eqref{Ls} \cite{Antola:2009wq,Alanne:2013dra}.
This pattern reflects the global symmetry (and its breaking)
of the strong sector of the NMWT fundamental
Lagrangian~\cite{Belyaev:2008yj,Appelquist:1999dq},

The vacuum expectation values (vev) of the $H$ and $M$ fields
break the EW symmetry and give
mass to both the SM and the TC states. The scalar sector can be recast in terms
of a type-I 2HDM~\cite{Alanne:2013dra}, while the vector mass eigenstates are
determined by diagonalizing the charged and neutral vector mass matrices given in
Appendix~\ref{VMassM}. The mass spectrum of the model, besides the SM particles,
features one heavy Higgs, $h'$, two mass degenerate pions, $a^0$ and
$h^\pm$, and finally two charged and two neutral vector bosons, $W'^\pm$, $W''^\pm$,
and $Z'$, $Z''$, respectively. The lightest charged vector boson, $W^\pm$, results from the mixing of the gauge field $\tilde{W}$ with composite vector fields that do not couple to SM fermions: consequentially the $W^\pm$ coupling to SM fermions is reduced. However, we checked that the Fermi coupling, $G_F$, determined by evaluating the tree level amplitude for the muon decay ($\mu^-\rightarrow \nu_\mu\bar{\nu}_e e^-$), respects the usual relation
\be
\sqrt{2} G_F=v_w^{-2}=(246\,{\textrm{GeV}})^{-2}\ .
\ee
 In the next section we perform a basic analysis
of the phenomenological viability of composite 2HDM at LHC.

\section{Viable Phenomenology and the LHC fit}
\label{couplingsfit}

We express the scalar mass parameters $m_M$ and $m_H$ in terms of the remaining parameters by minimizing the scalar potential in Eq.~\eqref{Ls} with respect to the vevs $f$ and $v$, which are determined by matching the experimental values of the EW scale, Eq.~\eqref{vfvw}, and the Higgs mass, 125~GeV. The free parameters of the model are therefore $\lambda_H$, $\lambda_M$, $y_{\rm TC}$, $c_1$, $c_2$ ,$c_3$, $c_4$,
from the scalar sector in Eq.~\eqref{Ls}, and $g_{\rm TC}$, $m_A$, $r_1$, and $r_2$, from the
vector sector in Eq.~\eqref{Lv}. To assess the viability of the model, we scan
the parameter space for data points that produce the observed SM mass spectrum,
and satisfy the lower bounds on the scalar and pseudoscalar masses
\be\label{mScon}
m_{h'}>600~{\rm GeV}\,;\ m_{a^0}\,,m_{h^\pm}>100~{\rm GeV}\,,
\ee
as well as the experimental bounds on the EW oblique parameters~\cite{Peskin:1991sw,Agashe:2014kda}
\be
S=0.00\pm0.08\,,\ T=0.05\pm 0.07\,,\ \rho(S,T)=90\%\,.
\ee
The region of parameter space that we scan, as described in~\cite{Alanne:2013dra},
is limited by potential stability and perturbativity in $\lambda_H$, $\lambda_M$, and $y_{\rm TC}$,
while the order of the coefficients $c_i$ is fixed by dimensional analysis~\cite{Manohar:1983md}.
For the remaining vector sector parameters we choose values in the region, natural for TC,
\be\label{Pcon}
500~{\rm GeV}<m_A<2500~{\rm GeV}\,,\ 2<g_{\rm TC}<6.5\,,\ r_1=-r_2={\cal O}(\frac{4m^2_A}{g^2_{\rm TC} f^2})\,,
\ee
where we impose the relation $r_1=-r_2$, which cancels the SM Higgs field coupling to the axial
combination $A^\mu_L+A^\mu_R$, to simplify the phenomenological analysis without
compromising the viability of the model.

Finally, we select the data points that satisfy the LHC constraints on the Higgs
couplings to $W$, $Z$, $\gamma$, $b$, and $\tau$, as done in ~\cite{Alanne:2013dra}:
for this purpose we calculate $\chi^2$ at each viable data point for the five Higgs coupling strengths, defined by
\be
\hat{\mu}_{jj}=\frac{\sigma_{pp\rightarrow h(X)} \textrm{Br}_{jj}}{\sigma_{pp\rightarrow h(X)}^{\textrm{SM}} \textrm{Br} ^{\textrm{SM}}_{jj}}\ ,\quad \textrm{Br}_{jj}=\frac{\Gamma_{h\rightarrow jj}}{\Gamma_h}\,,
\label{LHCb}\ee
where $X$ is a possible state produced in association with the light Higgs, and $jj$ a particle pair. The coupling strength values measured by both ATLAS and CMS~\cite{ATLASCMSfit} in inclusive processes are summarized in Table~\ref{datatable}.
\begin{table}[htb]
\begin{tabular}{|c||c|c|}
\hline
$jj$ & ATLAS & CMS \\
\hline
$ZZ$ & $\,1.52\pm 0.37\,$ & $\,1.04\pm 0.29\,$  \\
 $\gamma\gamma$ & $1.14\pm 0.26$ & $1.11\pm 0.24$  \\
$WW$ & $1.22\pm 0.22$ & $0.90\pm 0.22$  \\ 
$\tau\tau$ & $1.41 \pm 0.38$ & $0.88\pm 0.29$ \\
$bb$ & $0.62\pm 0.37$ & $0.81\pm 0.44$\\
\hline
\end{tabular}
\caption{Coupling strength experimental values determined by the ATLAS and CMS experiments.}
\label{datatable}
\end{table}

The Higgs couplings relevant for this analysis are those of SM particles and new resonances, contributing to leading order loop couplings, whose coupling coefficients are defined by
\bea\label{afVp}
{\cal{L}}_{\textrm{eff}} &\supset & a_V\frac{2m_W^2}{v_w}hW^+_\mu W^{-\mu}+a_V\frac{m_Z^2}{v_w}hZ_\mu Z^\mu
-a_f\sum_{\psi=t,b,\tau}\frac{m_\psi}{v_w}h\bar{\psi}\psi\nonumber \\
&+&a_{V'}\frac{2m^2_{W'}}{v_w}hW^{\prime +}_\mu W^{\prime -\mu}+a_{V''}\frac{2m^2_{W''}}{v_w}hW^{\prime\prime +}_\mu W^{\prime\prime -\mu}-a_h\frac{2 m_{h^\pm}^2}{v_w}h\,h^+ h^-
\,,
\eea
where all the fields in the equation above are physical eigenstates. The coefficients of the SM Higgs linear couplings to matter fields in Eq.~\eqref{afVp} can be expressed in terms of the Lagrangian parameters by
\bea\label{afVS}
a_h&=&\left[\left(c_{2 \beta }-c_{2 \rho }\right) \left(\left(c_2-c_4\lambda_H\right) c^{-1}_{\rho } s^{-1}_{\rho } \left(c_{\alpha +3 \beta }+c_{\alpha -\beta } c_{2 \beta } c_{2 \rho }\right)\right.\right.\nonumber\\
&+&\left.\left. 4 \left(c_2+c_4\lambda_H\right) c_{\beta } s_{\beta } \left(c_{\alpha } c_{\beta } t_{\rho }^{-2}+s_{\alpha } s_{\beta } t_{\rho }^2\right)\right)\right.\nonumber\\
   &-&\left.\left(c_{\alpha -\rho } s_{2 (\beta -\rho )}^2 s_{\beta +\rho } \lambda _H+c_{\alpha +\rho } s_{\beta -\rho } s_{2 (\beta +\rho )}^2 \lambda _M\right) c_{\rho }^{-2} s_{\rho }^{-2}/y_{TC}\right]\nonumber\\
&/&\left[4 \left(c_4\lambda_H s_{\beta -\rho }^2+\left(12 c_1+c_2\right) s_{\beta +\rho }^2\right) \right]\ ,\nonumber\\
a_f&=&\frac{c_{\alpha -\rho }}{s_{\beta -\rho }}\ ,\quad s_{\rho }=\sqrt{\frac{ 1 - c_3 y_{TC}}{2}}\ ,\quad c_{\rho }=\sqrt{\frac{ 1+c_3 y_{TC}}{2}}\,,
\eea
where $s_\alpha,c_\alpha,t_\alpha$ are shorthands for $\sin\alpha,\cos\alpha,\tan\alpha$, respectively, with $\alpha,\beta$ defined by the rotation matrices 
\be\label{rotM}
\begin{pmatrix} h \\ h^\prime \end{pmatrix} = \begin{pmatrix} c_\alpha & -s_\alpha \\ s_\alpha & c_\alpha \end{pmatrix} \begin{pmatrix} \phi \\ \varphi \end{pmatrix} ~, \
\begin{pmatrix} G^0 \\ a^0 \end{pmatrix} = \begin{pmatrix} s_\beta & c_\beta \\ c_\beta & -s_\beta \end{pmatrix} \begin{pmatrix} \pi_H^3 \\ \pi_M^3 \end{pmatrix} \,.
\ee
The coupling coefficients of the charged vector resonances in Eq.~\eqref{afVp} can be written in compact form by expanding at leading order correction in $\epsilon$ and $x$ as
\be\label{directa}
a_V=\eta_W s_{\beta - \alpha}\ ,\quad a_{V'}= \eta_{W'} s_{\beta - \alpha}\ ,\quad a_{V''}= \eta_{W''} s_{\beta - \alpha}\ ,
\ee
where
\bea\label{etaeq}
\eta_W&\cong& 1-\frac{\left[1+s^2 \left(3-\zeta\right)+2 s^4\right] x^2 \epsilon ^2}{\left(1+2 s^2\right)^2}\ ,\quad \eta_{W'}\cong\frac{2 \zeta s^2}{1+2 s^2}+\frac{\left[1+2 s^2\left(1-\zeta\right)\right]x^2 \epsilon ^2}{2 \left(1+2 s^2\right)^2}\ ,\nonumber\\ \eta_{W''}&\cong&\frac{x^2 \epsilon ^2}{2}\,,
\eea
with
\be\label{xesz}
s\equiv\frac{g_{\rm TC} f}{2 m_A}  \sqrt{r_1}\ ,\quad x\equiv \frac{g_L v_w}{2 m_A}\ ,\quad  \epsilon\ \equiv\frac{g_L}{g_{\rm TC}}\ ,\quad \zeta=s_{\beta-\alpha}^{-1}\frac{c_{\alpha+\rho}}{s_{\beta+\rho}}\ .
\ee
As one can see from the last of Eqs.~\eqref{etaeq}, the light Higgs coupling to $W''$ is negligible at leading order: this is a consequence of setting $r_1=-r_2$, Eqs.~\eqref{Pcon}, which makes the mixing term between the axial composite vector field and the SM gauge field, Eq.~\eqref{MWM}, small in the limit of small $\epsilon$. Given that the heavier vector resonances couple to SM fermions only through the mixing with the SM gauge fields, also the $W''$ couplings to SM fermions are small. Finally, the same statements are true also for $Z''$, given that setting $r_1=-r_2$ in Eq.~\eqref{MZM} makes the mixing term between axial and gauge vector fields small.

\begin{figure}[htb]
  \centering
    \includegraphics[width=.55\linewidth]{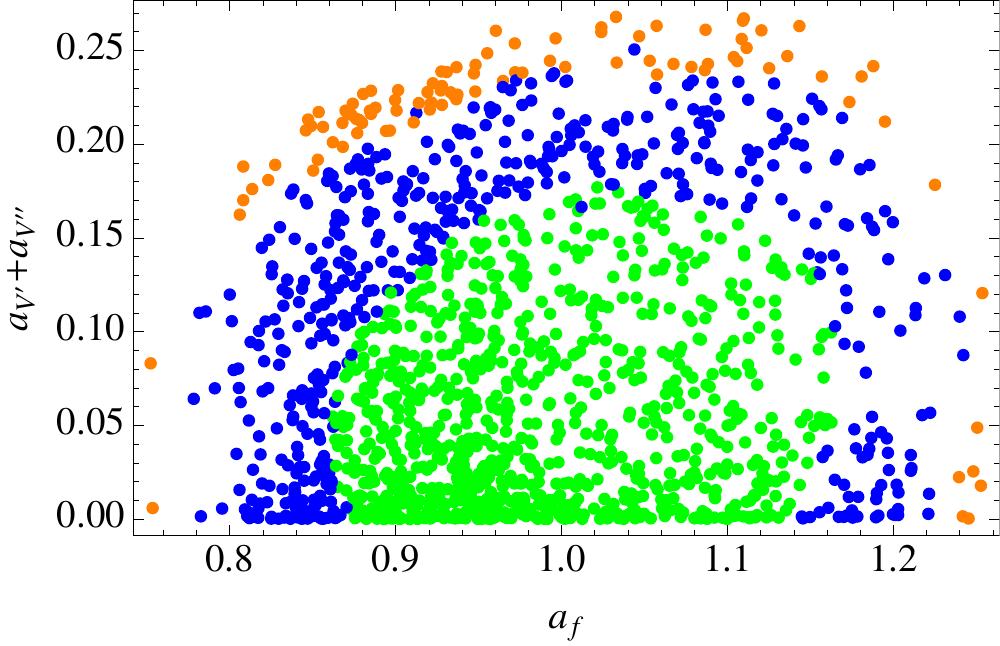}
  \caption{Viable points at 68\%,90\%,95\%CL shown in green, blue, orange in the
  plane of the coupling coefficients $a_f$ and $(a_{V'}+a_{V''})$ as defined in Eq.~\eqref{afVp}.}
  \label{afaVp}
\end{figure}

The charged non-SM particles in Eq.~\eqref{afVp} contribute only to the diphoton decay, while the decay rates of SM particles get rescaled by the square of the corresponding coupling coefficient.\footnote{All the relevant expressions, including those of the coupling coefficients in terms of the independent parameters of the model, are given in \cite{Alanne:2013dra}.} We then select the data points satisfying the 90\% confidence level (CL) constraint
\be\label{LHCfit}
P\left( \chi^2>\chi^2_{min} \right)>10\%
\ee
with 7 d.o.f., given that the number of observables is twelve while the effective free parameters is five. To see that the effective free parameters are just five one can notice that it is possible to fit simultaneously only five (three coupling strengths plus S and T) of the observables.\footnote{The effective free parameters for the Higgs linear couplings are just three: one for the fermions, one for the EW vector bosons, and one for the new physics contribution to the loop mediating the Higgs decay to diphoton.} 
Of the 10000 scanned data points satisfying perturbativity, potential stability, direct search constraints in Eqs.~\eqref{mScon}, and producing the observed SM mass spectrum, a total of 1381 points satisfy also the constraint in
Eq.~\eqref{LHCfit}: in Fig.~\ref{afaVp} these points are shown in green (68\%~CL)
and blue (90\%~CL), while those in orange are viable with 95\%~CL, in the
plane of the fermion and sum of the charged heavy vector resonances' coupling coefficients.\footnote{Notice that the contribution of the charged heavy vector resonances to the diphoton decay rate is proportional to the squared sum of their coupling coefficients, given that their masses are much heavier than the light Higgs mass.}
As shown in Fig.~\ref{afaVp}, the experimentally favored values of the coupling coefficients lie close to one for fermions and to zero for the heavy vector resonances, as expected given that the measured values of the $W$, $Z$, $\gamma$, $b$, $\tau$ coupling strengths are SM-like. On the other hand the heavy neutral vector resonances are not directly constrained by the Higgs coupling strengths fit.

In the next section we use the collection of 1381 data points viable at 90\%~CL to analyze the phenomenology
of both charged and neutral vector resonance production signals at the LHC.

\section{LHC Signals}
\label{LHCsigs}

\begin{figure}[h]
  \centering
    \includegraphics[width=.32\linewidth]{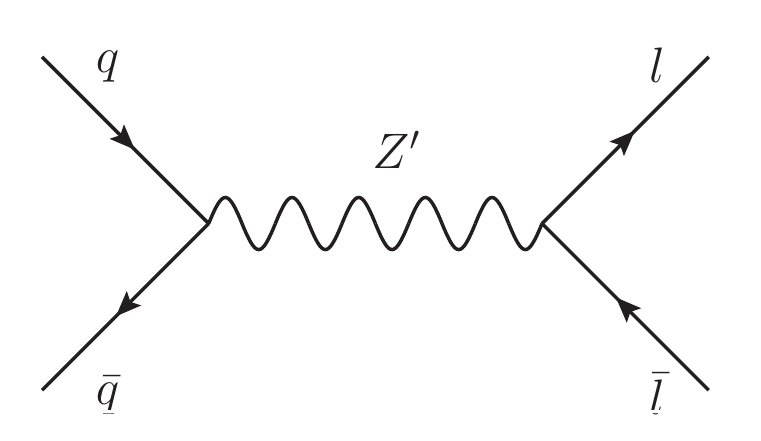}
    \includegraphics[width=.10\linewidth]{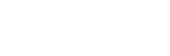}
    \includegraphics[width=.32\linewidth]{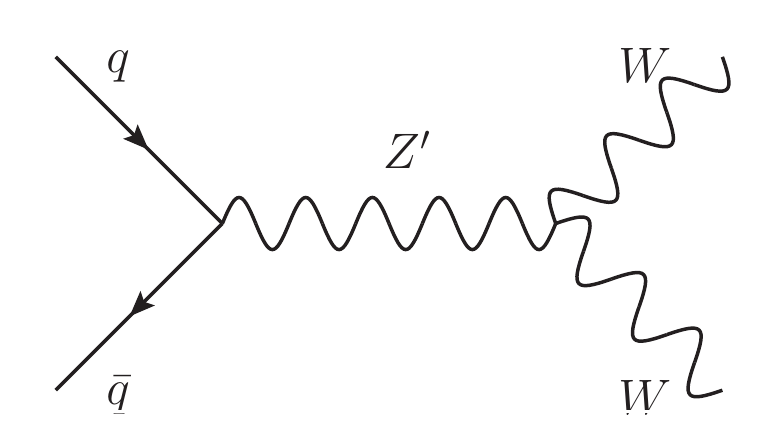}
  \caption{Feynman diagrams for the Drell-Yan production of a $Z'$ subsequently decaying to a dilepton or a diboson.}
  \label{DY}
\end{figure}

In order to compare the predictions of the composite 2HDM with the experimental
results from the LHC, we implemented the model, defined by Eqs.~(\ref{Ls},\ref{Lv})
as well as by the SM fermion and QCD sectors, in the Monte Carlo event generator
Madgraph~\cite{Alwall:2014hca} by using the Mathematica package
Feynrules~\cite{Christensen:2008py,Alloul:2013bka}. We have validated the implementation by
checking that the values of the relevant couplings of the vector boson mass
eigenstates evaluated with Madgraph at a sample data point matched the analytical
result for the same values of the input parameters. We have performed the collider
analysis at parton level, neglecting higher order effects such as parton showering,
intial and final state radiation, and detector resolution. The most constraining
final state turns out to be dimuon production for the neutral vector resonance, and charged lepton and neutrino for the charged vector resonance. These are electroweak processes
and therefore are not too sensitive to QCD effects or NLO corrections. Furthermore, the uncertainty
associated with our neglect of the detector resolution effects is mitigated by the fact that
the experimental resolution for lepton momentum is very high.
Thus we find it justifiable to perform the initial analysis at parton level.
The relevant production channels we present here are the Drell-Yan production of
a $Z'$ subsequently decaying to a dilepton or a diboson, expressed by the Feynman
diagrams in Fig.~\ref{DY}, and the production of a $W'$ decaying to a charged lepton and a neutrino.

\begin{figure}[htb]
  \centering
    \includegraphics[width=.60\linewidth]{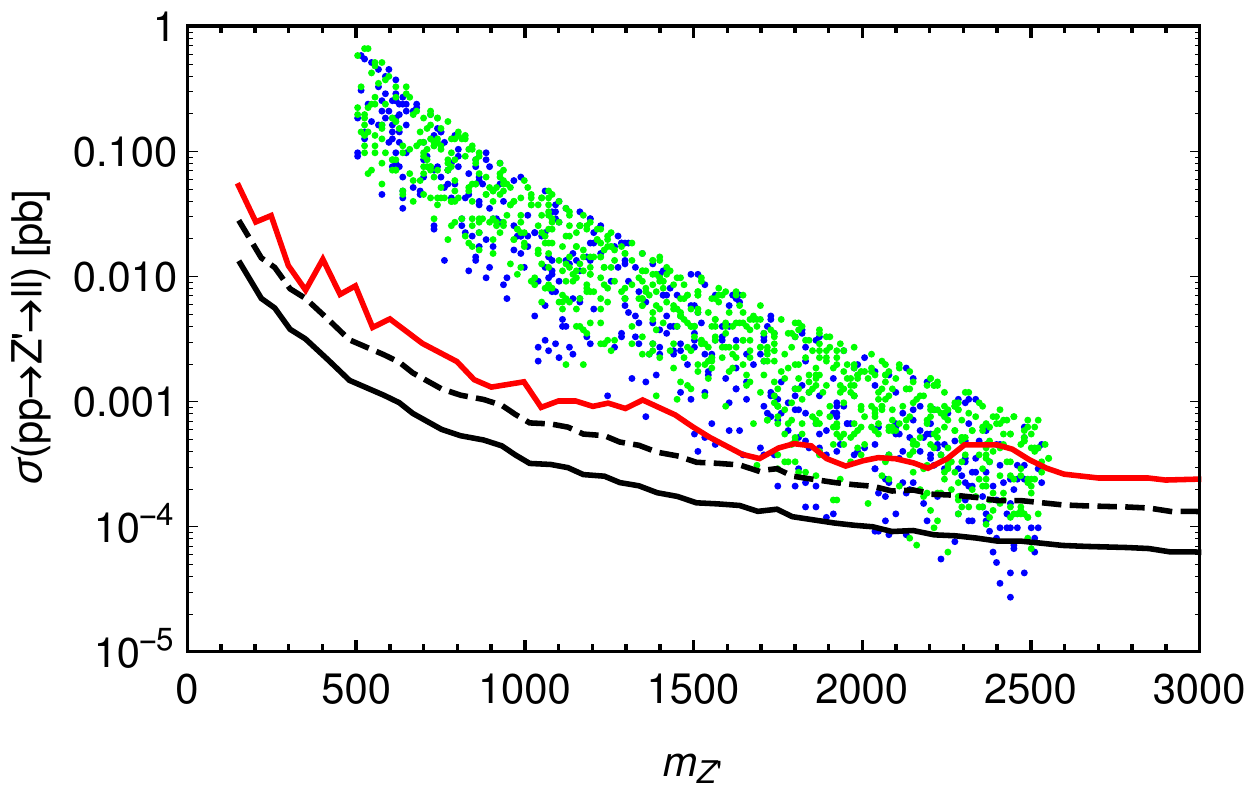}
  \caption{Upper bounds on the cross section for the Drell
  Yan production of a $Z'$ subsequently decaying to a dilepton: the solid red line is the observed exclusion limit from~\cite{ATLAS:2016cyf},
  the black dashed line shows the projected limit with 45~fb${}^{-1}$, and the black solid
  line shows the projected limit with 200~fb${}^{-1}$. The plot also shows the cross section for each
  of the 1381 viable data points evaluated with MadGraph. The color code of the data points
  corresponds to that of Fig.~\ref{afaVp}.}
\label{dilepton}
\end{figure}

The ATLAS dilepton search~\cite{ATLAS:2016cyf} looks for two opposite sign isolated
charged leptons within the pseudorapidity window  \mbox{$|\eta|<2.5$}. The corresponding
production cross section evaluated with Madgraph at each of the 1381 collected data
points is shown in Fig.~\ref{dilepton} together with the upper limit (red solid line)
from~\cite{ATLAS:2016cyf} and the projected limit with 45~fb${}^{-1}$ (black dashed line)
and 200~fb${}^{-1}$ (black solid line). The color code of the data points corresponds
to that of Fig.~\ref{afaVp}. We find that out of the scanned data points, 86\% are
already ruled out by the dilepton search, 94\% can be ruled out with the currently
existing data of 45~fb${}^{-1}$, and 99\% with the projected integrated luminosity
of 200~fb${}^{-1}$ for the current run. As explained above, we have limited the mass
of the vector resonance to \mbox{$m_{Z'}\leq 2.5$ TeV}, in order to stay safely below
the TC confinement scale $\Lambda_{\rm TC}\sim 3$~TeV.
Increasing further the composite vector boson masses, while generally allowed by naive dimensional analysis, would be less desirable based on naturalness arguments, which disfavor a large hierarchy between the confinement and the electroweak scales.

\begin{figure}[h]
\centering
    \includegraphics[width=.60\linewidth]{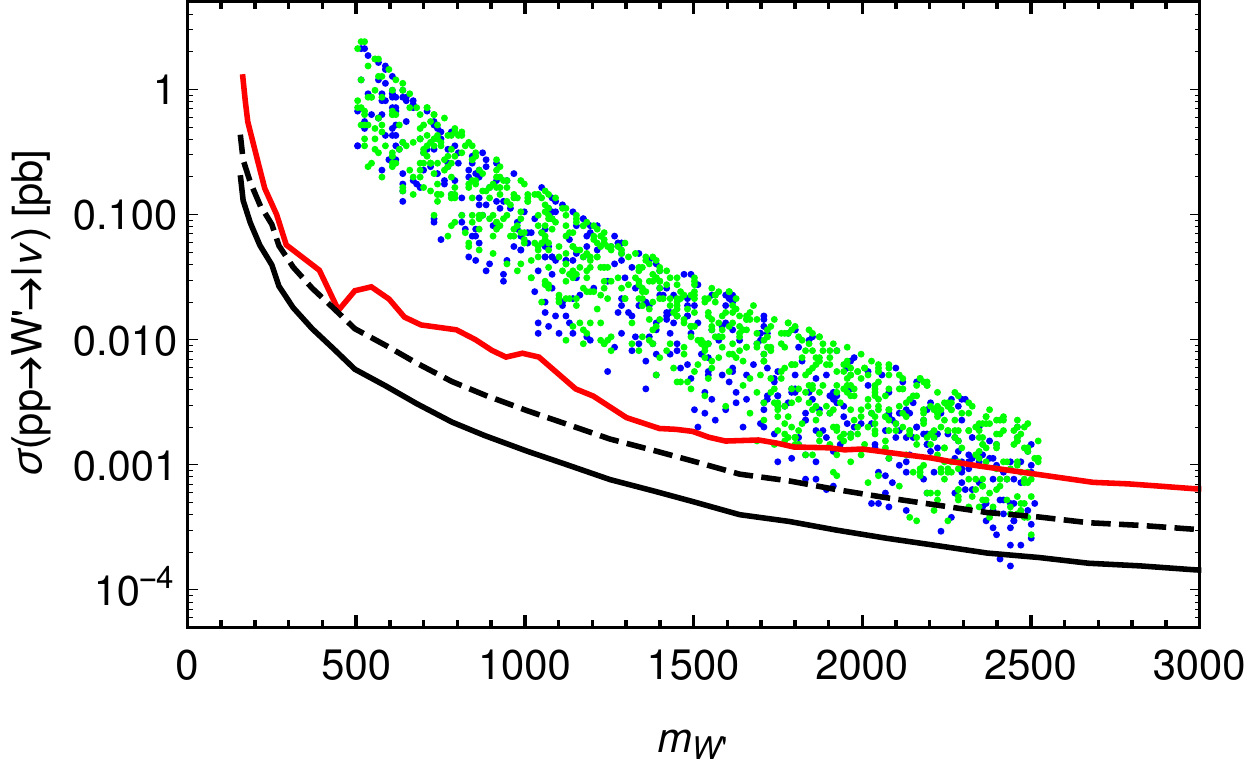}
  \caption{Upper bounds on the cross section for the production of a $W'$ subsequently decaying to a charged lepton and a neutrino: the solid red line is the observed exclusion limit from~\cite{ATLAS:2016ecs},
  the black dashed line shows the projected limit with 45~fb${}^{-1}$, and the black solid
  line shows the projected limit with 200~fb${}^{-1}$. The plot also shows the cross section for each
  of the 1381 viable data points evaluated with MadGraph. The color code of the data points
  corresponds to that of Fig.~\ref{afaVp}.}
\label{Wprime}
\end{figure}

The most relevant constraint for the charged vector resonance $W'$ is given by the search for a charged lepton and missing energy~\cite{ATLAS:2016ecs}. This search selects events with a single muon with transverse momentum \mbox{$p_T > 55$ GeV} or a single electron with \mbox{$p_T > 65$ GeV} in the pseudorapidity window $|\eta|<2.5$ for the muons and $|\eta|<2.47$ for the electrons. Additionally, the events must contain significant missing energy, \mbox{$\slashed{E}_T>55$ GeV} in the muon channel and \mbox{$\slashed{E}_T > 65$ GeV} in the electron channel, and the transverse mass of the charged lepton plus neutrino system must be above 110 GeV in the muon channel and above 130 GeV in the electron channel. The corresponding cross section as a function of the $W'$ mass for the data points is shown in figure \ref{Wprime}, together with the constraint from~\cite{ATLAS:2016ecs}. The expected exclusion limits for 45~fb${}^{-1}$ and 200~fb${}^{-1}$ are shown by the black dashed and solid lines, respectively. We find that 90\% of the scanned data points are already ruled out by the $W'$ search, 98\% can be ruled out with the data set of 45~fb${}^{-1}$ and nearly all (99.8\%) of the scanned parameter space points are within reach with 200~fb${}^{-1}$.

\begin{figure}[h]
  \centering
    \includegraphics[width=.60\linewidth]{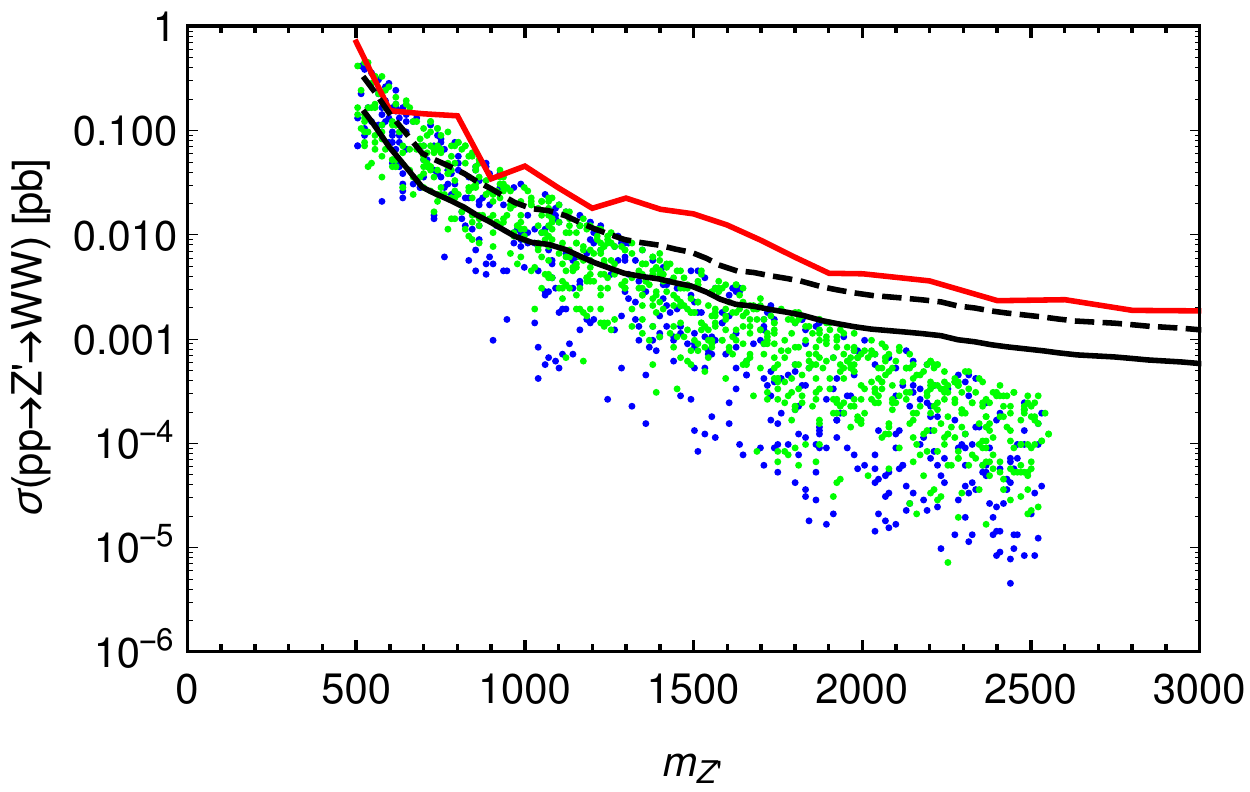}
  \caption{Upper bounds on the cross section for the Drell
  Yan production of a $Z'$ subsequently decaying to $WW$: the solid red line is the observed exclusion limit from~\cite{ATLAS:2016cwq},
  the black dashed line shows the projected limit with 45~fb${}^{-1}$, and the black solid
  line shows the projected limit with 200~fb${}^{-1}$. The plot also shows the cross section for each
  of the 1381 viable data points evaluated with MadGraph. The color code of the data points
  corresponds to that of Fig.~\ref{afaVp}.}
\label{diboson}
\end{figure}

The limit on the diboson channel is less tight, as can be seen from Fig.~\ref{diboson},
and most of the scanned data points are below the current limit~\cite{ATLAS:2016cwq},
shown by the red solid line. The expected exclusion limits for 45~fb${}^{-1}$ (black dashed line)
and for 200~fb${}^{-1}$ (black solid line) reach larger portions of the data points,
but all of these points are already ruled out by the dilepton search. From the color
code of the data points in figures~\ref{dilepton}, \ref{Wprime} and~\ref{diboson} one can see that the lower $\chi^2$ points shown in green tend to have a higher production
cross section. This feature seems more pronounced in the diboson channel, but is
present also in the dilepton and lepton plus neutrino channels.

\section{Conclusions and outlook}
\label{checkout}

The latest LHC direct searches of heavy vector
resonances~\cite{ATLAS:2016cyf,ATLAS:2016ecs,ATLAS:2016yqq,ATLAS:2016cwq,ATLAS:2016npe,CMS:2016pfl,CMS:2016abv} have an energy
reach comparable to the TC scale ($\sim 3$~TeV), and their constraints are therefore
relevant for TC models. In this paper we briefly reviewed a template TC model for
which EW symmetry breaking, triggered by the new TC strong interaction between EW
doublet technifermions, is transmitted to SM fermions via a scalar field coupling
the TC matter sector to the SM one. The model features two (partially)
composite Higgs
bosons and several new heavy vector bosons, which are fully composite states and
represent a clear signature common to all TC models~\cite{Andersen:2011yj}. We tested
the viability of this model first of all by performing a fit of the lighter Higgs
scalar couplings to SM vector bosons, bottom quarks, and tau leptons. We performed
the goodness of fit analysis by scanning the model's parameter space for data points
producing a viable SM particle mass spectrum, and selecting those points that satisfy
at 90\%~CL the experimental constraints on those couplings as well as the lower bound
on the mass of a heavy Higgs scalar. The selected data set consists of 1381 viable
data points. We then implemented the model in the event generator
Madgraph~\cite{Alwall:2014hca} and calculated, for each viable data point,
the cross section at the parton level for the dilepton and diboson channels
of Drell-Yan production of heavy neutral vector bosons, and lepton plus neutrino channel for the production of a charged vector boson at the LHC.
By comparing these results with the latest LHC constraints we showed that
a major portion of the otherwise viable data points is already excluded by
the direct searches of heavy vector resonances in the dilepton and lepton plus neutrino channels, while
almost the entire parameter space we have considered will be tested by
the end of LHC Run II in 2022.
On the other hand the experimental constraints on the diboson channel are less tight as they are able to rule
out only a smaller portion of the selected data set. These results show that
the LHC experiments have the potential to discover signatures of TC
in direct resonant production channels. Alternatively, if
no new heavy vector boson resonances are discovered, very stringent constraints
will be imposed on the TC framework of the type we have considered here, as a significant portion of parameter space naturally selected by naive dimensional analysis would be ruled out.

Heavy vector boson direct searches at the ATLAS and CMS experiments can in
principle complement flavor experiments carried out at LHCb, whose 2015 data
show large deviations from the SM predictions in flavor violating observables
which might well be explained by a new heavy neutral vector
particle~\cite{Altmannshofer:2015sma,Descotes-Genon:2015uva,Hurth:2016fbr,Gauld:2013qba}. Flavor
violating interaction terms in extended TC models are a natural by-product of
fermion mass terms, and therefore would be a well motivated extension of the present
template TC model.

\acknowledgments
The financial support from the Academy of Finland under grant 267842 is gratefully acknowledged.
We thank Dr. Tuomas Hapola for collaborating in the initial stages of this work.

\appendix

\section{Vector Mass Matrices}
\label{VMassM}
The neutral vector squared mass matrix, obtained from Eq.~\eqref{Lv}, is
\be\label{MZM}
{\cal M}_{Z}^2=\left(
\begin{array}{cccc}
 \epsilon ^2 m_A^2 t_\theta^2+m_{\tilde{W}}^2 \left(z_1+1\right) t_\theta^2 & -m_{\tilde{W}}^2 \left(z_2+1\right) t_\theta & -\frac{\epsilon  t_\theta
   m_A^2}{\sqrt{2}}-\frac{m_{\tilde{W}}^2 \left(z_1-z_2\right) t_\theta}{\sqrt{2} \epsilon } & \frac{\epsilon  t_\theta m_A^2}{\sqrt{2}}+\frac{m_{\tilde{W}}^2
   \left(z_1+z_2\right) t_\theta}{\sqrt{2} \epsilon } \\
 -m_{\tilde{W}}^2 \left(z_2+1\right) t_\theta & \epsilon ^2 m_A^2+m_{\tilde{W}}^2 \left(z_1+1\right) & -\frac{\epsilon  m_A^2}{\sqrt{2}}-\frac{m_{\tilde{W}}^2
   \left(z_1-z_2\right)}{\sqrt{2} \epsilon } & -\frac{\epsilon  m_A^2}{\sqrt{2}}-\frac{m_{\tilde{W}}^2 \left(z_1+z_2\right)}{\sqrt{2} \epsilon } \\
 -\frac{\epsilon  t_\theta m_A^2}{\sqrt{2}}-\frac{m_{\tilde{W}}^2 \left(z_1-z_2\right) t_\theta}{\sqrt{2} \epsilon } & -\frac{\epsilon
   m_A^2}{\sqrt{2}}-\frac{m_{\tilde{W}}^2 \left(z_1-z_2\right)}{\sqrt{2} \epsilon } & m_A^2+\frac{m_{\tilde{W}}^2 \left(z_1-z_2\right)}{\epsilon ^2} & 0 \\
 \frac{\epsilon  t_\theta m_A^2}{\sqrt{2}}+\frac{m_{\tilde{W}}^2 \left(z_1+z_2\right) t_\theta}{\sqrt{2} \epsilon } & -\frac{\epsilon
   m_A^2}{\sqrt{2}}-\frac{m_{\tilde{W}}^2 \left(z_1+z_2\right)}{\sqrt{2} \epsilon } & 0 & m_A^2+\frac{m_{\tilde{W}}^2 \left(z_1+z_2\right)}{\epsilon ^2}
\end{array}
\right)\,,
\ee
and the charged vector squared mass matrix is
\be\label{MWM}
{\cal M}_{W}^2=\left(
\begin{array}{ccc}
 \epsilon ^2 m_A^2+m_{\tilde{W}}^2 \left(z_1+1\right) & -\frac{\epsilon  m_A^2}{\sqrt{2}}-\frac{m_{\tilde{W}}^2 \left(z_1-z_2\right)}{\sqrt{2} \epsilon } & -\frac{\epsilon
   m_A^2}{\sqrt{2}}-\frac{m_{\tilde{W}}^2 \left(z_1+z_2\right)}{\sqrt{2} \epsilon } \\
 -\frac{\epsilon  m_A^2}{\sqrt{2}}-\frac{m_{\tilde{W}}^2 \left(z_1-z_2\right)}{\sqrt{2} \epsilon } & m_A^2+\frac{m_{\tilde{W}}^2 \left(z_1-z_2\right)}{\epsilon ^2} & 0 \\
 -\frac{\epsilon  m_A^2}{\sqrt{2}}-\frac{m_{\tilde{W}}^2 \left(z_1+z_2\right)}{\sqrt{2} \epsilon } & 0 & m_A^2+\frac{m_{\tilde{W}}^2 \left(z_1+z_2\right)}{\epsilon ^2}
\end{array}
\right)\,,
\ee
where
\be
m_{\tilde{W}}=\frac{g_L v_w}{2}\ ,\quad t_\theta=\frac{g_Y}{g_L}\ , \quad z_1=\frac{f^2}{v_w^2} r_1\ , \quad z_2=\frac{f^2}{v_w^2} r_2\,,
\ee
while the remaining quantities are defined in Eqs.~\eqref{xesz}.

\end{document}